\documentstyle[12pt,epsf]{article}
\renewcommand{\baselinestretch}{1.30}
\begin{document}
%
\begin{flushright}
  December, 1999 \ \ \\
  OU-HEP-338 \ \\
  hep-ph/9912500
\end{flushright}
\vspace{0mm}
\begin{center}
\large{Spin and Orbital Angular Momentum Distribution Functions
of the Nucleon}
\end{center}
\vspace{0mm}
\begin{center}
M.~Wakamatsu\footnote{Email \ : \ wakamatu@miho.rcnp.osaka-u.ac.jp}
\end{center}
\vspace{-4mm}
\begin{center}
Department of Physics, Faculty of Science, \\
Osaka University, \\
Toyonaka, Osaka 560, JAPAN
\end{center}
\vspace{0mm}
\begin{center}
T.~Watabe\footnote{Email \ : \ watabe@rcnp.osaka-u.ac.jp}
\end{center}
\vspace{-4mm}
\begin{center}
Research Center for Nuclear Physics (RCNP), \\
Osaka University, \\
Ibaraki, Osaka 567, JAPAN
\end{center}

\vspace{8mm}

\vspace{10mm}
\begin{center}
\small{{\bf Abstract}}
\end{center}
\vspace{-1mm}
\begin{center}
\begin{minipage}{15.5cm}
\renewcommand{\baselinestretch}{1.0}
\small
\ \ \ A theoretical prediction is given for the spin and orbital angular
momentum distribution functions of the nucleon within the framework
of an effective quark model of QCD, i.e. the chiral quark soliton
model. An outstanding feature of the model is that it predicts fairly
small quark spin fraction of the nucleon $\Delta \Sigma \simeq 0.35$,
which in turn dictates that the remaining $65 \,\%$ of the nucleon spin
is carried by the orbital angular momentum of quarks and antiquarks
at the model energy scale of $Q^2 \simeq 0.3 \,\mbox{GeV}^2$.
This large orbital angular momentum necessarily affects the scenario
of scale dependence of the nucleon spin contents in a drastic way.

\normalsize
\end{minipage}
\end{center}
\renewcommand{\baselinestretch}{2.0}

\newpage
\section{Introduction}

The recent renewal of interest in polarized lepton-hadron and
hadron-hadron deep-inelastic scatterings greatly owes to the
so-called ``nucleon spin crisis'' aroused by the EMC
experiment \cite{EMC88}.
According to the latest SMC analysis \cite{SMC98},
the intrinsic quark spin carries only about $20 \,\%$ of
the total nucleon spin.
The widely-believed resolution of this puzzle is based on an
assumption of large gluon polarization \cite{AR88}.
In fact, the recent next-to-leading-order (NLO) analyses of
the polarized deep-inelastic-scattering (DIS) data for
$g_1 (x,Q^2)$ appear to indicate a positive value for the
gluon polarization \cite{GS96}.
Nevertheless, the situation is far from conclusive, since the
result of these analyses is rather sensitive to the input shapes
of the polarized distributions as well as the uncertainties of
the available polarized DIS data.

Although somewhat nonstandard, there is another scenario that
also leads to qualitative resolution of the nucleon spin problem.
This scenario was first introduced on the basis of the Skyrme
model \cite {BEK88}.
Later, the idea was developed into a more quantitative one
within the framework of the chiral quark soliton model (CQSM)
\cite{W90},\cite{WY91}.
The CQSM is an effective quark model of QCD, which was first
introduced by Diakonov and Petrov, based on the instanton-liquid
picture of the QCD vacuum \cite{DP86},\cite{DPP88}.
Being in conformity with Witten's idea of large 
$N_c$ QCD \cite{Witten84}, it shares many features in common with
the Skyrme model \cite{Skyrme61},\cite{ANW83} at least in the
ideal limit of $N_c = \infty$. (For review, see \cite{RevCQSM}.)
At the subleading order of $1 / N_c$ expansion, however, several
crucial differences have been found to exist between
them \cite{WY91},\cite{Waka96},\cite{WW93}.
Among others, most important would
be the following two. The first is the $1 / N_c$ correction (or
the first order rotational correction in the collective angular
velocity $\Omega$) to the isovector axial-vector coupling constant
of the nucleon, which definitely exists in the CQSM but is entirely
missing in the Skyrme model \cite{WW93}.
(In our opinion, this solves the so-called ``$g_A$ problem''
in the latter \cite{Waka96}.)
The second concerns the subject of our central interest here,
i.e. the predictions for the quark spin fraction $\Delta \Sigma$
of the nucleon. As was first noticed by Brodsky et al. \cite{BEK88},
$\Delta \Sigma$ identically vanishes in the naive Skyrme model.
At first sight, this appeared to be a favorable feature of the
model since it looks consistent with the result of
the EMC measurement \cite{EMC88}.
As we have argued repeatedly \cite{WY91},\cite{Waka96}, however,
this outstanding prediction of the Skyrme model should be taken with
care, because it just arises from the fact that this model cannot
correctly describe the physics at the subleading order of
$1 / N_c$ expansion. In fact, the presence of the
next-to-leading order contribution to $\Delta \Sigma$ is again
a distinguishable feature of the CQSM as compared with the
naive Skyrme model \cite{WY91}.
Nonetheless, qualitative similarity between
the two models still survives
and the prediction of the CQSM for $\Delta \Sigma$ remains to be
fairly small, i.e. $\Delta \Sigma \simeq 0.35$, as compared with
other effective models of the nucleon like the nonrelativistic
quark model or the MIT bag model. What carries the rest of the
nucleon spin ? Within the CQSM, the answer to this question is quite
simple. It must be the orbital angular momentum of quarks and/or
antiquarks, since it is an effective theory with the effective quark
degrees of freedom only \cite{WY91}.

A natural question is whether the effective quark degrees of freedom
are enough to describe the relevant physics. One might suspect that
the gluon polarization or its orbital angular momentum may play
important roles even at the low energy hadronic scale.
There have been some attempts to introduce the explicit gluon
degrees of freedom into the CQSM in reference to the underlying
instanton-liquid picture of the QCD vacuum \cite{DPW96},\cite{DMW99}.
Unfortunately, the situation is far from conclusive yet.
Here we simply assumes that the
explicit gluonic degrees of freedom are not crucial at least for
the polarized distribution functions at the low renormalization point,
which is taken to be the starting energy of the DGLAP evolution
equation. That this is not a meaningless assumption can, for instance,
be convinced from our recent analysis of the longitudinally polarized
distribution functions of the nucleon and the deuteron in comparison
with the recent EMC and SMC data \cite{WK99},\cite{WW99}.
This analysis starts with the predictions of the CQSM for the
longitudinally polarized distribution functions for quarks and
antiquarks, which are taken to be input distributions given at the
model energy scale of $Q_{ini}^2 = 0.3 \,\mbox{GeV}$. The polarized
distribution functions for the gluon are simply assumed to be zero
at this initial energy scale. After solving the standard evolution
equation at the NLO, the predictions of the model were then compared
with the recent EMC and SMC data for $g_1^p (x,Q^2), g_1^n (x,Q^2)$,
and $g_1^d (x,Q^2)$ given at $Q^2 = 5 \,\mbox{GeV}^2$ \cite{SMC98}.
Despite the rather drastic assumption of zero gluon polarization
at the hadronic scale, the theory was shown to reproduce all the
qualitatively noticeable features of the experimental data.

Encouraged by this success, now we push forward our attempts to
clarify the spin contents of the nucleon. That is, the main
purpose of the present study is to give a theoretical prediction
for the orbital angular momentum distribution functions for quarks
and antiquarks in the nucleon. After combining those with the
previously-obtained quark spin distribution functions, we can
investigate the scale dependence of the full nucleon spin
contents by using the recently-derived evolution equation
at the leading order (LO) \cite{HS98},\cite{HK99}.
(We however recall fact that there is a criticism against
this LO evolution equation for the orbital angular momentum
distributions \cite{HJL99}.)
Although our analysis is bound to
one basic assumption that the gluon polarization as well as
the gluon orbital angular momentum play no significant role
at least at the hadronic scale of $Q^2 \leq 0.3 \,\mbox{GeV}$,
it is expected to provide us with quite a unique scenario for
the scale dependence of the nucleon spin content, while
keeping satisfactory agreement with the available high-energy
data for $g_1^p (x,Q^2), g_1^n (x,Q^2)$, and $g_1^d (x,Q^2)$.

We shall explain in the next section the theoretical framework
which our calculation of the orbital angular momentum distribution
functions for quarks and antiquarks are based on.
Next, we discuss the scale dependence of the nucleon spin
contents including the orbital angular momentum based on the given
input distributions at the low renormalization point.
Finally, we summarize our findings in the last section.

\section{The theoretical framework}

How to treat the orbital angular momentum has been one of the most
controversial theoretical problems in hadron spin physics.
It is well known that the most natural definition of angular
momentum is not gauge invariant \cite{JM90}.
Recently, Bashinsky and Jaffe
proposed a gauge-invariant formulation of the angular momentum
for quarks and gluons which reduces to natural definition in
a particular gauge, i.e. the light-cone gauge
$A^+ (x) = 0$ \cite{BJ98}.
It was also suggested \cite{MHS99} that this gauge invariant
formulation may give an extended support to the recently derived
LO evolution equation for the orbital angular momentum
distributions in the light-cone gauge \cite{HS98},\cite{HK99}. 
According to the authors of \cite{HJL99}, however,
nonlocal operators with dependence on spatial coordinates have
not been seen in factorization of hard forward scattering
processes, and in particular inclusive deep-inelastic scatterings
do not depend on such types of operators. 
Accodingly, the evolution of the orbital angular momentum 
distributions would become extremely complicated in the 
light-cone gauge contrary to the results of
\cite{HS98},\cite{HK99}. On the other hand,
there is another definition of orbital angular momentum
distributions, which maintains gauge invariance at the cost of
losing full Lorentz covariance \cite{HJL99}.
This latter definition has its
meaning only in a limited class of coordinates in which the nucleon
has a definite helicity. An advantage of the second definition is
that it can be extracted from the polarized and unpolarized quark
distributions and the off-forward distributions $E(x)$ so that it
can in principle be measured.
What is not fully clarified yet is the relation between these two
definitions of angular momentum distributions.
In sum, the proper definition of the orbital angular momentum
distributions as well as the corresponding evolution equation in
QCD is still an unsettled issue being subject to a debate. 
One should therefore take the following investigation of 
the orbital angular momentum distribution within an effective
theory of QCD bearing all these affairs in mind.

Since the CQSM at the present level of approximation contains no
gluonic degrees of freedom at least explicitly, it seems natural
to start with the following naive definition of the quark orbital
angular momentum distribution function :
\begin{equation}
 q_L (x) \ = \ \frac{1}{\sqrt{2} \,p^+} \,\int_{-\infty}^\infty \,
 \frac{d \lambda}{2 \,\pi} \,\,e^{\,i \,\lambda \,x} \,
 < P S_3 \,| \,\psi_+^\dagger (0) \,i \, 
 ( \,x^1 \partial^2 - x^2 \partial^1 \,) \,\psi_+ (\lambda n)
 \,| \,P S_3 > ,
\end{equation}
Here $p^\mu$ and $n^\mu$ are two light-like (null) vectors,
having the properties
\begin{equation}
 p^- = 0, \ \ n^+ = 0, \ \ p^2 = n^2 = 0, \ \ p \cdot n = 1,
\end{equation}
while $\psi_+$ is a component of the quark field $\psi$
defined through the decomposition
\begin{equation}
 \psi \ = \ (P_+ + P_-) \,\psi \ = \ \psi_+ \ + \ \psi_- ,
\end{equation}
by the projection operators $P_{\pm} = \frac{1}{2} 
\gamma^{\mp} \gamma^{\pm}$
with $\gamma^{\pm} = (1 / \sqrt{2})(\gamma^0 \pm \gamma^3)$.
Extending the definition of distribution function $q_L(x)$ to
interval $-1 \leq x \leq 1$, the relevant antiquark 
distributions are given as 
\begin{equation}
 \bar{q}_L(x) \ = \ q_L(-x), \ \ \ \ \ \ \ \ (0 < x < 1) .
\end{equation}
Naturally, from more general viewpoint of underlying color
gauge theory, i.e. QCD, the naive definition (1) is not gauge
invariant and holds only in the light-cone gauge. As was already
explained, this definition of the quark orbital angular momentum
is not free from problems, especially when we are to use it as a 
initial distribution of scale evolution. We expect that this
is not so serious for our investigation below, since our
main concern there is the qualitative behavior of the quark
orbital angular momentum distribution at the model scale, which
naturally has a strong correlation with the quark spin distribution
at the same scale. On the other hand, the study of its scale
dependence should be taken as semi-qualitative one.

The basis of our theoretical analysis is the following path 
integral representation of the nucleon matrix element of 
bilocal quark operator \cite{DPPPW97} :
\begin{eqnarray}
 < N (\mbox{\boldmath $P$}) \,| \,\psi^+(0) \,O \,\psi(z) \,| \,
 N(\mbox{\boldmath $P$}) >
 &=& \frac{1}{Z} \int d^3 x \,d^3 y \,\,e^{-i \mbox{\boldmath $P$} 
 \cdot \mbox{\boldmath $x$}}
 e^{i \mbox{\boldmath $P$} \cdot \mbox{\boldmath $y$}} 
 \int{\cal D} \mbox{\boldmath $\pi$}
 \int {\cal D} \psi \,{\cal D} \psi^\dagger \nonumber \\
 &\times& \!\!
 J_N (\frac{T}{2}, \mbox{\boldmath $x$}) \cdot \psi^\dagger (0)
 \,O \,\psi(z) \cdot J_N^\dagger (-\frac{T}{2}, \mbox{\boldmath $y$})
 \,\,e^{\,i \int d^3 x \,{\cal L}_{CQM}} , \ \ 
\end{eqnarray}
where
\begin{equation}
 {\cal L}_{CQM} \ = \ \bar{\psi} \,(i \not\!\partial - M 
 e^{i \gamma_5 \mbox{\boldmath $\tau$} \cdot 
 \mbox{\boldmath $\pi$}(x) / f_{\pi}}) \,\psi ,
\end{equation}
is the basic lagrangian of the CQSM, while
\begin{equation}
  J_N (x) \ = \ \frac{1}{N_c \,!} \,\,
  \epsilon^{\alpha_1, \cdots, \alpha_{N_c}} \,\,
  \Gamma^{f_1,\cdots,f_{N_c}}_{J J_3, T T_3} \,\,
  \psi_{\alpha_1 f_1} (x) \cdots \psi_{\alpha_{N_c} f_{N_c}} (x),
\end{equation}
is a composite operator carrying the quantum numbers $J J_3, T T_3$
(spin, isospin) of the nucleon, where $\alpha_i$ is the color index,
while $\Gamma^{f_1,\cdots,f_{N_c}}_{J J_3, T T_3}$ is a symmetric
matrix in spin-flavor indices $f_i$.
We start with a stationary 
pion field configuration of hedgehog shape :
\begin{equation}
 \mbox{\boldmath $\pi$}(x) = f_{\pi} \,\hat{\mbox{\boldmath $r$}} \,F(r) .
\end{equation}
Next we carry out a path integral over $\mbox{\boldmath $\pi$}(x)$
in a saddle point approximation by taking care of two
zero-energy modes, i.e. the translational zero-modes and
rotational zero-modes. Under the assumption of ``slow rotation''
as compared with the intrinsic quark motion, the answer can be 
obtained in a perturbative series in $\Omega$, which can also be
regarded as a $1 / N_c$ expansion. The first nonvanishing 
contribution to $q_L(x)$ arises at the $O(\Omega^1)$ term of this 
expansion, since the leading $O(\Omega^0)$ term vanishes 
identically due to the hedgehog symmetry. According to the general 
formalism derived in \cite{WK99}, the answer is given in the
following form :
\begin{equation}
 \Delta q_L(x) \ = \ \Delta q_L^{ \{ A, B \} }(x) \ + \ 
 \Delta q_L^C(x) ,
\end{equation}
where
\begin{eqnarray}
 \Delta q_L^{ \{ A , B \} } \!\!\! &=&  \ \ <J_3>_{p \uparrow} \cdot 
 M_N \,\frac{N_c}{I} \sum_{m = all, n \leq 0}
 \frac{1}{E_m - E_n} < n \,| \,\tau_3 \,|\, m >
 < m \,| \,(1 + \gamma^0 \gamma^3) L_3 \,\delta_n \,|\, n > 
 \nonumber \\
 \!\!\! &=& \!\! - <J_3>_{p \uparrow} \cdot M_N \,
 \frac{N_c}{I} \sum_{m = all, n > 0} \frac{1}{E_m - E_n}
 < n \,| \,\tau_3 \,| m > < m \,| \,
 (1 + \gamma^0 \gamma^3) L_3 \,\delta_n \,|\, n > ,
 \ \ \ \ \ \ 
\end{eqnarray}
and
\begin{eqnarray}
 \Delta q_L^C(x) &=& \ \ \,<J_3>_{p \uparrow} \,\cdot \,\frac{d}{d x}
 \,\,\frac{N_c}{2 I} \,\,\sum_{n \leq 0}
 < n \,| \,\tau_3 \,(1 + \gamma^0 \gamma^3) \,L_3 \,\delta_n \,|\, n > 
 \nonumber \\
 &=& \!\! - \,<J_3>_{p \uparrow} \,\cdot \,\frac{d}{d x} \,\,
 \frac{N_c}{2 I} \,\,\sum_{n > 0}
 < n \,| \,\tau_3 \,(1 + \gamma^0 \gamma^3) \,L_3 \,\delta_n \,| \,n > ,
\end{eqnarray}
with $L_3 \equiv - \,i \,(x^1 \partial^2 - x^2 \partial^1)$ and 
$\delta_n \equiv \delta (x M_N - E_n - p_3)$, while
${\langle {\cal O} \rangle}_{p \uparrow}$ denotes a matrix element
of a collective space operator ${\cal O}$ with respect to the proton
in the spin up state along the $z$-axis, i.e.
\begin{equation}
  {\langle {\cal O} \rangle}_{p \uparrow} \ = \ 
  \int \,\Psi^{(1/2)}_{(1/2) (1/2)} [\xi_A] \,{\cal O} \,
  \Psi^{(1/2)}_{(1/2) (1/2)} [\xi_A] \,d \xi_A \ = \ 
  \langle p, S_3 = 1/2 \,| \,{\cal O} \,| \, p, S_3 = 1/2 \rangle .
\end{equation}
Here, $|\, m >$ and $E_m$ are the eigenstates and the associated 
eigenenergies of the static hamiltonian 
$H = - \,i \,\mbox{\boldmath $\alpha$} \cdot 
\mbox{\boldmath $\nabla$} + \beta M \,e^{\,i \,\gamma_5 \,
\mbox{\boldmath $\tau$} \cdot \hat{\mbox{\boldmath $r$}} \,F(r)}$
with the hedgehog background.
Note that both of $\Delta q_L^{ \{ A, B \} }(x)$ and $q_L^C(x)$ are 
represented in two equivalent forms, i.e. in the occupied form
$(n \leq 0)$ convenient for the numerical calculation for
$x > 0$ and in the nonoccupied form $(n > 0)$ convenient
for $x < 0$.

In the actual numerical calculation, the expression of any
physical quantity is divided into two pieces, i.e. the contribution
of what we call the valence quark level ( it is the lowest energy
eigenstate of the static hamiltonian $H$, which emerges from the
positive energy continuum) and that of the Dirac sea quarks
(the latter is also called the vacuum polarization
contribution) as explained in \cite{WK99}.
Since the latter contains ultraviolet divergences, it must be
regularized. Here we use the so-called Pauli-Villars
regularization scheme \cite{DPPPW97}. The regulator mass
$M_{PV}$ of this cutoff scheme is determined uniquely from the
physical requirement that the effective meson action derived
from (6) gives correct normalization for the pion kinetic term.
Using the value of $M = 375 \,\mbox{MeV}$, which is favored from
the phenomenology of nucleon low energy observables, this
condition gives $M_{PV} \simeq 562 \,\mbox{MeV}$. Since we are
to use these values of $M$ and $M_{PV}$, there is {\it no free
parameter} additionally introduced in the calculation of
distribution functions.

\section{Numerical Results and Discussion}

We first show in Fig.1 the theoretical predictions for the orbital
angular momentum distribution functions of quarks and antiquarks
as well as the isosinglet quark polarization. (The latter was
already given in \cite{WK99}, but it is shown here again for
emphasizing how the $x$-dependences of the two distributions
are different.)
Here the distribution with negative $x$ are to be interpreted
as the antiquark distribution according to the rule (4) and the
similar relation for $\Delta \bar{u}(x) + \Delta \bar{d}(x)$.
The long-dashed and dash-dotted curves respectively stand for the
contribution of the discrete valence level and that of the negative
energy Dirac sea, while their sum is shown by 
the solid curves. As already noticed in \cite{WK99}, the contribution
of Dirac sea is not so significant for $\Delta u(x) + \Delta d(x)$.
The situation is quite different for the orbital angular momentum
distribution $q_L (x)$. One clearly sees that the contribution of the 
polarized Dirac vacuum being peaked around $x \simeq 0$ is
dominating over that of the discrete valence level.
This can also be convinced from the first moment defined by
\begin{equation}
 L_q \ = \ \int^1_{-1} \,q_L (x) \,d x \ = \ \int^1_0 \,
 [\,q_L (x) + \bar{q}_L (x) \,] \,d x .
\end{equation}

\begin{figure}[htbp] \centering
\epsfxsize=16cm 
\epsfbox{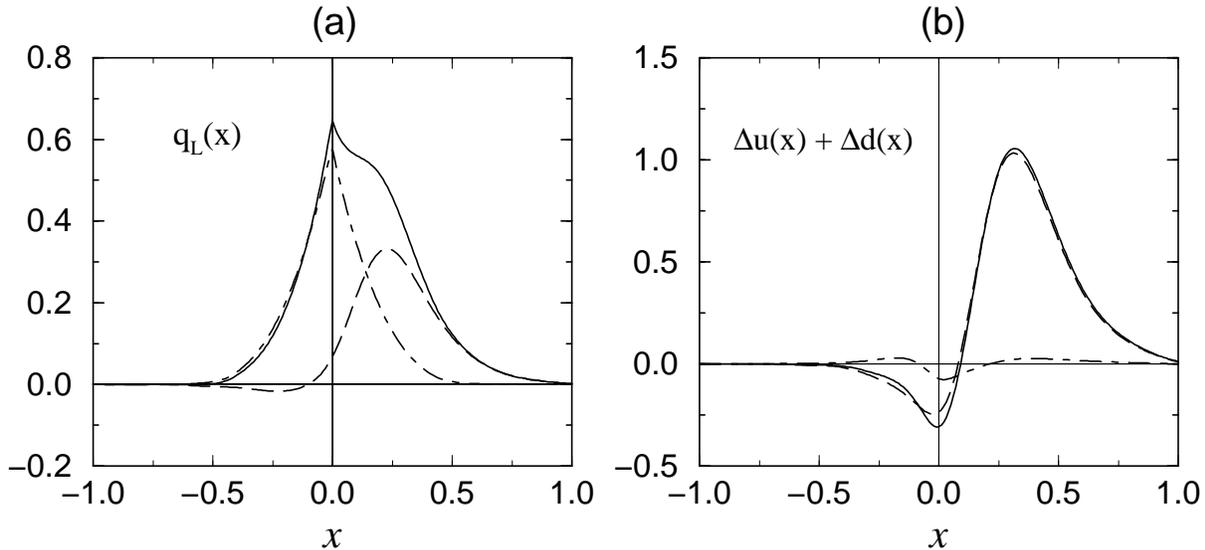}
\renewcommand{\baselinestretch}{1.00}
\caption{(a) The theoretical predictions of the CQSM for the quark
and antiquark orbital angular momentum distribution functions
$q_L(x)$ and (b) the isosinglet quark polarization
$\Delta u(x) + \Delta d(x)$.
The long-dashed and dash-dotted curves respectively stand for
the contributions of the discrete valence level and that of the
Dirac continuum in the self-consistent hedgehog background,
whereas their sums are shown by the solid curves. The distributions
with negative $x$ are to be interpreted as the antiquark
distributions.} 
\renewcommand{\baselinestretch}{1.20}
\end{figure}

\ \\
The contributions of the Dirac continuum and of the discrete
valence level to this moment are respectively 0.195 and 0.130, 
which shows that the former dominates over the latter.
The sum of these two numbers gives $L_q \simeq 0.325$, which means 
that about $65 \,\%$ of the nucleon spin is carried by the orbital
angular momentum of quarks and antiquarks at the energy scale
of the model \cite{WY91},\cite{WK99}.

\vspace{2mm}
\begin{table}[h]
\begin{center}
\renewcommand{\baselinestretch}{1.2}
\caption{The separate contributions of quarks and antiquarks to
the first moment $\Delta \Sigma$ and $L_q$ at the scale of the model.}
\renewcommand{\baselinestretch}{1.38}
\vspace{6mm}
\begin{tabular}{cccc} \hline\hline
 & \ \ \ quark \ \ \ & \ \ \ antiquark \ \ \ & \ \ \ total \ \ \ \\
\hline \hline
\ \ \ $\Delta \Sigma$ \ \ \ & 0.397 & - \,0.047 \, & 0.350 \\ \hline
\ \ \ $L_q$ \ \ \ & 0.229 & 0.096 & 0.325 \\ \hline
\ \ \ $\frac{1}{2} \,\Delta \Sigma \,+ \,L_q$ \ \ \ &
0.427 & 0.073 & 0.500 \\ \hline\hline
\end{tabular}
\end{center}
\end{table}
We can also know the separate contributions of quarks and antiquarks
to the total orbital angular momentum. They are shown in table 1 
together with the corresponding separation for the quark spin
fraction of the nucleon, or the flavor-singlet quark polarization
$\Delta \Sigma$. One sees that the flavor-singlet polarization 
of antiquark is negative but fairly small in magnitude. (Note
however that this does not necessarily mean that the polarizations
of all the antiquarks are small.
It has been shown \cite{WW99} that the CQSM predicts fairly large
isospin asymmetry of the longitudinally polarized antiquark
distribution in the nucleon, i.e. $\Delta \bar{d} (x) - \Delta 
\bar{u} (x) < 0$.)
On the contrary, a sizable amount ($\sim \,30 \,\%$)
of the total orbital angular momentum is seen to be carried by
antiquarks with soft (small $x$) component.

At this point, we want to make a short comment on the recent
lattice QCD calculation of quark orbital angular
momentum \cite{MDLMM99}.
Aside from the fact that it is based on the so-called quenched
approximation, one should also notice the fact that this is
an indirect calculation of the quark orbital angular momentum
$L_q$ as well as the total gluon angular momentum $J_g$.
They first extract the total quark angular momentum $J_q$
from the calculation of the quark energy-momentum tensor form
factors. Further combining the previous lattice
calculation \cite{DLF95} of the quark spin content
$\Delta \Sigma$, they obtain $L_q$, thereby extracting $J_g$
from the total nucleon spin sum rule. This dictates that
the extracted values of $L_q$ and $J_g$ would be sensitive to the
uncertainties of $J_q$ and $\Delta \Sigma$ obtained in the
numerical simulation. The main conclusion of this analysis is
that about $25 \%$ of the nucleon spin originates from the quark
intrinsic spin, while about $35 \%$ comes from the quark orbital
angular momentum, which in turn means that the remaining $40 \%$
of the nucleon spin is due to the glue.
If this is confirmed by more precise calculation in the future,
the present investigation done with neglect of the
explicit gluonic degrees of freedom needs a considerable amendment.
At least, some of the quark contributions obtained within the
QCSM must be redistributed to either or both of the gluon spin
and orbital angular momentum. It is interesting to see that
scaling the CQSM predictions $\langle \frac{1}{2} \,\Delta \Sigma
\rangle \simeq 35 \%$ and $\langle L_q \rangle \simeq 65 \%$
by the factor of 0.6, we would obtain $\langle \frac{1}{2} \,
\Delta \Sigma \rangle \simeq 21 \%$ and $\langle L_q \rangle
\simeq 39 \%$, which is rather close to the corresponding numbers
$25 \%$ and $35 \%$ obtained in the above lattice QCD study.
An interesting common feature in both theories is the
dominance of the quark orbital angular momentum over the
intrinsic quark spin fraction.

Now we are in a position to investigate the scale dependence of
the full nucleon spin contents together with the corresponding 
distribution functions. As already mentioned, we assume that 
the gluon polarization and the gluon orbital angular momentum 
are both zero at the starting energy scale of  
$Q^2 = Q_{ini}^2$, i.e.
\begin{eqnarray}
 \Delta \Sigma (Q_{ini}^2) \ = \ 0.350, \ \ 
 \Delta g (Q_{ini}^2) \ = \ 0, \ \
 L_q (Q_{ini}^2) \ = \ 0.325, \ \
 L_g (Q_{ini}^2) \ = \ 0,
\end{eqnarray}
with the normalization
\begin{equation}
 \frac{1}{2} \,\Delta \Sigma (Q^2) \ + \ \Delta g(Q^2) 
 \ + \ L_q(Q^2) \ + \ L_g(Q^2) \ = \ \frac{1}{2} \,\,.
\end{equation}
The question here is how to choose $Q_{ini}^2$. In our previous
analysis of the longitudinally polarized structure functions
$g_1^p (x, Q^2), g_1^n (x, Q^2)$ and $g_1^d (x, Q^2)$ based 
on the evolution equation at the next-to-leading-order (NLO), a good 
agreement with the available high-energy data has been obtained with 
the choice $Q_{ini}^2 = 0.30 \,\mbox{GeV}^2$ \cite{WK99},\cite{WW99}.
Unfortunately, the evolution equation including the orbital angular
momentum distributions is known only at the leading
order (LO) \cite{JTH96}. Expecting that the use of a 
little smaller value of $Q_{ini}^2$ would compensate the defect of
using this lower-order evolution equation, we proceed as follows. 
That is, we first solve the evolution equation for 
$\Delta \Sigma (Q^2)$ and $\Delta g (Q^2)$ at the NLO ( in the
$\overline{MS}$ scheme) by setting 
$Q_{ini}^2 = 0.30 \,\mbox{GeV}^2$ \cite{FP82},\cite{GRV90}.
Next, we solve the evolution equation 
for the full nucleon spin contents at the LO with use of a little 
smaller $Q_{ini}^2$, i.e. $Q_{ini}^2 = 0.23 \,\mbox{GeV}^2$,
which is determined so as to roughly reproduce 
the $Q^2$-evolution of the gluon polarization  
$\Delta g (Q^2)$ obtained by solving the NLO evolution with 
$Q_{ini}^2 = 0.30 \,\mbox{GeV}^2$.

\begin{figure}[htbp] \centering
\epsfxsize=16cm 
\epsfbox{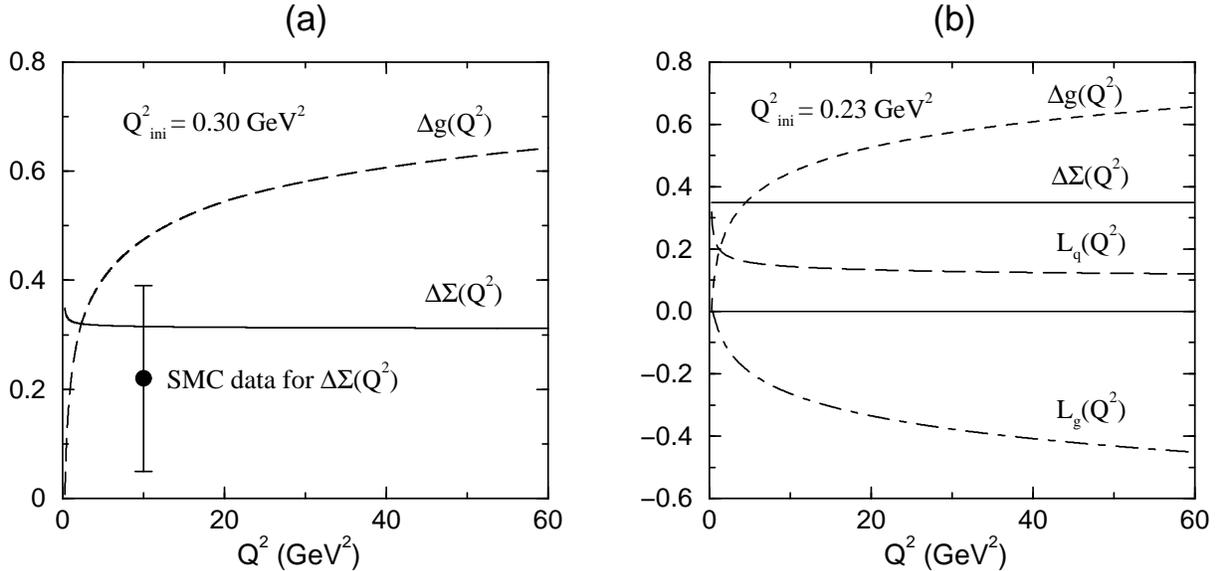}
\renewcommand{\baselinestretch}{1.00}
\caption{(a) The NLO evolutions of the flavor-singlet quark and
gluon polarizations. The experimental data given at $Q^2 = 10 \,
\mbox{GeV}^2$ corresponds to the NLO analysis by the SMC
group \cite{SMC98}. (b) The LO evolution of the full nucleon
spin contents.}
\renewcommand{\baselinestretch}{1.20}
\end{figure}

We first show in Fig.2(a) the NLO evolution for the quark and gluon
polarization in comparison with the latest SMC data for 
$\Delta \Sigma (Q^2)$.
One can say that the prediction of the CQSM for $\Delta \Sigma$ is
qualitatively consistent with the recent NLO analysis of the SMC
group given at $Q^2 = 10 \,\mbox{GeV}^2$ \cite{SMC98}.
One also sees that the
gluon polarization rapidly grows with increasing $Q^2$, even if
we have assumed $\Delta g = 0$ at the initial energy scale of 
$Q_{ini}^2 = 0.30 \,\mbox{GeV}^2$.

Shown in Fig.2(b) is the LO evolution of the full nucleon spin 
contents including the orbital angular momentum.
They are calculated by using the known analytical
solutions \cite{JTH96} :
\begin{eqnarray}
  \Delta \Sigma (Q^2) &=& \ \ \Delta \Sigma (Q_{ini}^2) \ \ = \ \ 
  \mbox{constant} ,\\
  \Delta g (Q^2) &=& - \,\frac{12}{33-2 \,n_f} \,
  \Delta \Sigma (Q_{ini}^2) \ + \ \frac{t}{t_0} \,
  \left( \,\Delta g (Q_{ini}^2) + 
  \frac{12}{33-2 \,n_f} \,\Delta \Sigma (Q_{ini}^2) 
  \, \right) , \\
  L_q (Q^2) &=& - \,\frac{1}{2} \,\Delta \Sigma (Q_{ini}^2)
  \ + \ \frac{1}{2} \,\frac{3 \,n_f}{16 + 3 \,n_f} \nonumber \\
  &\,& \hspace{10mm} + \, 
  {\left( \frac{t}{t_0} 
  \right)}^{- \frac{2 \,(16 + 3 \,n_f)}{3 \,(33 - 2 \,n_f)}}
  \,\left( \,L_q (Q_{ini}^2) + \frac{1}{2} \,
  \Delta \Sigma (Q_{ini}^2)
  - \frac{1}{2} \,\frac{3 \,n_f}{16 + 3 \,n_f} \right), \\
  L_g (Q^2) &=& - \,\Delta g (Q_{ini}^2)
  \ + \ \frac{1}{2} \,\frac{16}{16 + 3 n_f} \nonumber \\
  &\,& \hspace{10mm} + \,
  {\left( \frac{t}{t_0} 
  \right)}^{- \frac{2 \,(16 + 3 \,n_f)}{3 \,(33 - 2 \,n_f)}}
  \,\left( \,L_g (Q_{ini}^2) + \Delta g (Q_{ini}^2)
  - \frac{1}{2} \,\frac{16}{16 + 3 n_f} \right) ,
\end{eqnarray}
with $t \equiv \ln Q^2 / \Lambda_{QCD}^2$, $t_0 \equiv 
\ln Q_{ini}^2 / \Lambda_{QCD}^2$ and $n_f = 3$. 
The starting energy of this evolution is taken to be 
$Q_{ini}^2 = 0.23 \,\mbox{GeV}^2$, as mentioned above.
As is widely-known, the flavor-singlet quark polarization
$\Delta \Sigma$ is scale independent at the LO.
Also relatively established is
that the gluon polarization $\Delta g(Q^2)$ shows a logarithmic
growth with increasing $Q^2$, while it is largely compensated by
the similar decrease of the gluon orbital angular momentum
$L_q (Q^2)$. In fact, it was pointed out by Ji et al. \cite{JTH96}
that, as $Q^2 \rightarrow \infty$, the sum of these two
moments approaches its asymptotic value given by
\begin{equation}
  J_q \ \equiv \ \Delta g \ + \ L_g \ \longrightarrow \ 
  \frac{1}{2} \,\frac{16}{16 + 3 \,n_f} ,
\end{equation}
which indicates (with $n_f = 3$) that about $64 \%$ of the nucleon
spin is carried by the gluon fields in the asymptotic energy.

On the other hand, the evolution of quark orbital angular momentum is
strongly dependent on the initial condition given at the low
renormalization point. Since the scale-dependent factor
\begin{equation}
  {\left( \frac{t}{t_0} \right)}^{- \frac{2 \,
  (16 + 3 \,n_f)}{3 \,(33 - 2 \,n_f)}} ,
\end{equation}
is a decreasing function of $Q^2$ (for $33 - 2 n_f > 0$), we conclude
that $L_q (Q^2)$ is a decreasing function of $Q^2$ as far as
\begin{equation}
  L_q (Q_{ini}^2) \ + \ \frac{1}{2} \,\Delta \Sigma (Q_{ini}^2)
  \ - \ \frac{1}{2} \,\frac{3 \,n_f}{16 + 3 \,n_f} ,
\end{equation}
is positive, while it is a increasing function of $Q^2$ if the
same quantity above is negative. In the case of the CQSM, it gives
$L_q (Q_{ini}^2) + \frac{1}{2} \,\Delta \Sigma (Q_{ini}^2) 
- \frac{1}{2} \,\frac{3 \,n_f}{16 + 3 n_f} = 0.325 + 0.175
- 0.18 > 0$ at $Q^2 = Q_{ini}^2$, so that it is consistent with
the fact that $L_q (Q^2)$ shown in Fig.3 is a decreasing function
of $Q^2$. The $Q^2$ dependence of $L_q (Q^2)$ looks strong only in
the relatively lower $Q^2$ region, and beyond $Q^2 \simeq 5 \,
\mbox{GeV}^2$ it slowly approaches its asymptotic value given by
\begin{equation}
  L_q (Q^2) \ \longrightarrow \ - \,\frac{1}{2} \,
  \Delta \Sigma (Q_{ini}^2) \ + \ \frac{1}{2} \,
  \frac{3 \,n_f}{16 + 3 n_f} .
\end{equation}
Substituting our model predictions $\Delta \Sigma (Q_{ini}^2) = 0.35$
with $n_f = 3$, this gives $L_q \longrightarrow 0.005$, which denotes
that the quarks barely carry orbital angular momentum at the very
high energy scale. Note, however, that the precise fraction of the
quark orbital angular momentum in the asymptotic domain depends on
the delicate cancellation of the two terms in (23).

It appears that the observation above contradicts one of the
conclusions drawn in the recent paper by Scopetta and
Vento \cite{SV99}.
They concluded that the quark orbital angular momentum can be
important at large $Q^2$, due to evolution, if it is not
negligible at the scale of the hadronic model.
It seems that this conclusion is mainly drawn from Fig.5(b)
in \cite{SV99}, which shows that $L_q (Q^2)$ is an increasing
function of $Q^2$.
This increasing behavior of $L_q (Q^2)$ appears to be inconsistent
with the $Q^2$-dependence of the quark orbital angular momentum
distributions illustrated in Fig.4(a)
of the same paper. In fact, the initial conditions of their ``D-model''
are given by $\Delta \Sigma (Q_{ini}^2) \simeq 0.4$, 
$\Delta g (Q_{ini}^2) \simeq 0.1$, $L_q (Q_{ini}^2) \simeq 0.145$,
and $L_g (Q_{ini}^2) \simeq 0.055$.
Since this gives $L_q (Q_{ini}^2) + \frac{1}{2} \,\Delta
\Sigma (Q_{ini}^2) - \frac{1}{2} \,\frac{3 \,n_f}{16 + 3 n_f}
= 0.165 > 0$, $L_q (Q^2)$ must be a decreasing function of $Q^2$
if our argument above is correct.
Probably, the cause of discrepancy resides in an error in
eqs.(17), (18) of \cite{SV99}.
The $Q^2$-dependent factor $b^{-38/81}$
(with the definition $b = \alpha_S (Q^2) / \alpha_S (Q_{ini}^2)$)
in (17) and (18) should be replaced by $b^{+50/81}$ for $n_f = 3$.
(Note that $b^{-38/81}$ is an increasing function of $Q^2$, while
$b^{+50/81}$ is a decreasing function.)

Next, we show in Fig.3 the theoretical prediction of the CQSM for the
spin and orbital angular momentum distribution functions and their LO
evolutions. Here, we have used the LO evolution code provided by
Martin et al. \cite{MHS99}.
Here, the solid and long-dashed curves respectively represent
$x \,\Delta \Sigma (x,Q^2)$ and $x \,L_q (x,Q^2)$, whereas the dashed
and dash-dotted curves stand for $x \,\Delta g(x,Q^2)$ and
$x \,L_g (x,Q^2)$. At the initial energy scale of $Q_{ini}^2 = 0.23
\,\mbox{GeV}$, we set $\Delta g(x) = L_g (x) = 0$.
On the other hand, $\Delta \Sigma (x)$ and $L_q (x)$ at the same
scale are obtained as
\begin{eqnarray}
  \Delta \Sigma (x) &=& \left[ \Delta u(x) + \Delta d(x) \right]
  \ + \ \left[ \Delta \bar{u} (x) + \Delta \bar{d} (x) \right], \\
  L_q (x) &=& q_L (x) \ + \ \bar{q}_L (x),
\end{eqnarray}
from the predictions of the CQSM shown in Fig.1.
The logarithmically increasing
behavior of $\Delta g(x,Q^2)$ and the logarithmically decreasing
behavior of $L_g(x,Q^2)$ are just what has been pointed out
in several previous papers \cite{MHS99},\cite{SV99}.

\begin{figure}[htbp] \centering
\epsfxsize=16cm 
\epsfbox{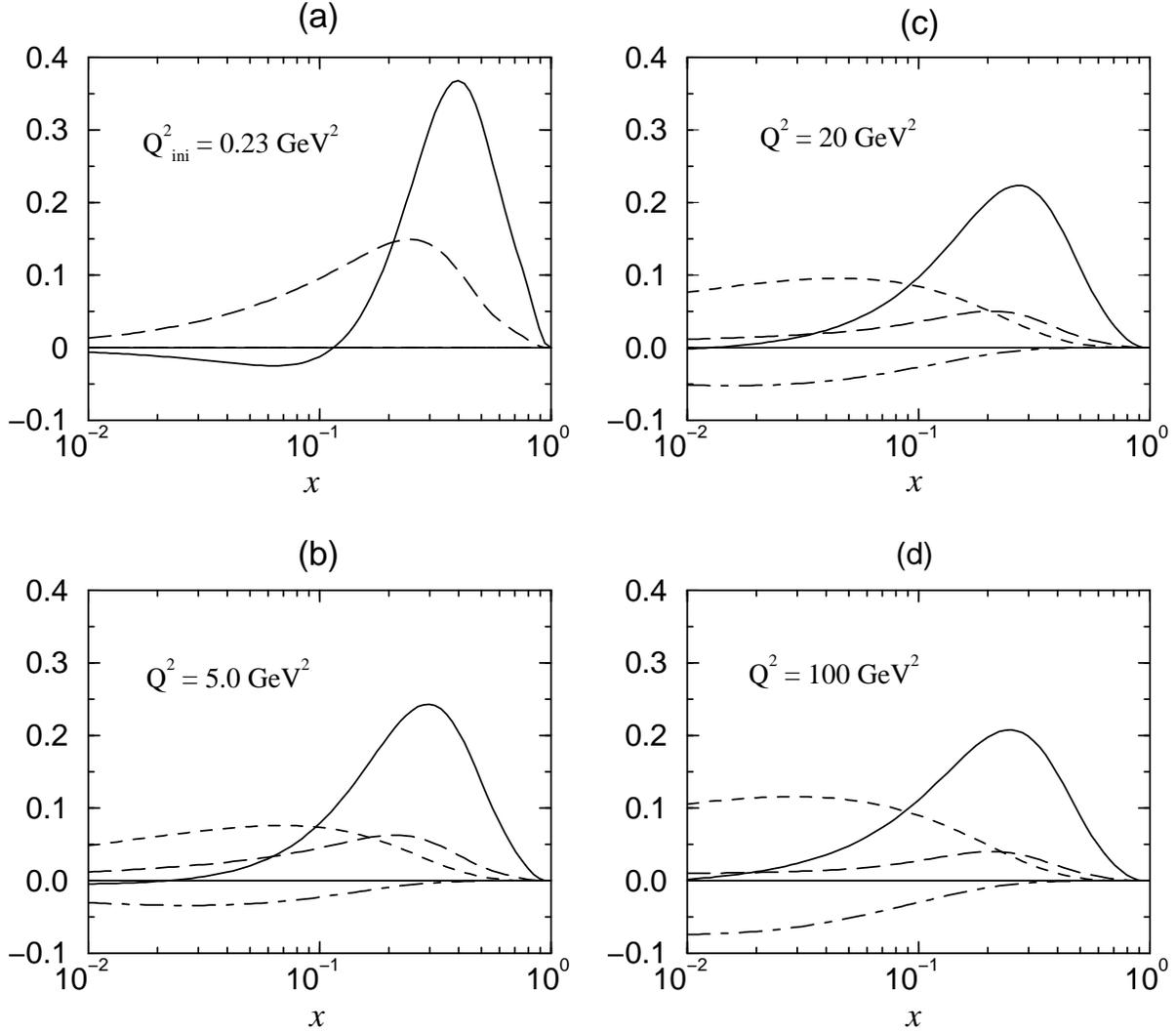}
\renewcommand{\baselinestretch}{1.00}
\caption{The LO evolutions of the spin and orbital angular
momentum distribution functions for quarks and gluons.
the solid and long-dashed curves respectively represent
$x \,\Delta \Sigma (x,Q^2)$ and $x \,L_q (x,Q^2)$, whereas the dashed and
dash-dotted curves stand for $x \,\Delta g(x,Q^2)$ and $x \,L_g (x,Q^2)$.}
\renewcommand{\baselinestretch}{1.20}
\end{figure}

\ \\
As pointed out in \cite{MHS99},
$L_g (x,Q^2)$ at high energy scale is rather insensitive to the input
distributions for orbital angular momentum at the starting energy,
and it is basically determined by the polarized quark singlet and
gluon distributions. On the other hand, the behavior of
$L_q(x,Q^2)$ at the moderate value of $Q^2$ is strongly
dependent on the input orbital angular momentum distribution.
In fact, Fig.4 shows that $L_q(x,Q^2)$ remains positive
at all the energy scales, although its magnitude gradually
decreases as $Q^2$ increases. We point out that this behavior
of $L_q(x,Q^2)$ is consistent with the scale dependence of the
corresponding first moment $L_q(Q^2)$ shown in Fig.2(b).

Clearly, the origin of this unique behavior of $L_q(x,Q^2)$
and $L_q(Q^2)$ can be traced back to the extraordinary large
and positive orbital angular momentum for quarks and
antiquarks at
the hadronic scale predicted by the CQSM.
The results above may be compared with several previous
investigations of the scale dependence of the orbital
angular momentum distributions, the input distributions of which
are prepared in a somewhat arbitrary way.
For instance, two different scenario have been
investigated in \cite{MHS99}. The first scenario called the GRSV
standard scenario has the property that $\frac{1}{2} \,\Delta
\Sigma (Q_{ini}^2) + \Delta g (Q_{ini}^2) \simeq 0.475$
with very small orbital angular momentum for quarks and gluons,
i.e. $L_q (Q_{ini}^2) + L_g (Q_{ini}^2) \simeq 0.015$.
The second rather extreme scenario assumes very large gluon
polarization such that $\Delta g (x,Q_{ini}) = g (x,Q_{ini}^2)$,
which in turn dictates large and negative orbital angular
momentum such that $L_q (Q_{ini}^2) + L_g (Q_{ini}^2) = - \,
0.43$. A common feature of these scenarios is that the evolved
$L_q (x,Q_{ini}^2)$ has large and negative values in wide
range of $x$, which is quite different from the above-mentioned
result of the CQSM shown in Fig.4.
Unfortunately, any of these quite different
scenarios cannot be excluded on the basis of our present knowledge,
since all of these reproduces the available data for the
longitudinally polarized deep-inelastic structure functions for
the nucleon and the deuteron at least qualitatively.
Highly desirable is some direct experimental information for
any of $\Delta g(x,Q^2)$, $L_q (x,Q^2)$ and $L_g (x,Q^2)$.

\section{Conclusions}

An outstanding feature of the CQSM is that it can explain all
the qualitatively noticeable features of the recent high-energy
deep-inelastic-scattering observables including NMC data for
$F^p_2 (x) - F^n_2 (x)$, $F^n_2 (x) / F^p_2 (x)$ \cite{NMC91},
the Hermes \cite{HERMES} and NuSea \cite{E866} data for
$\bar{d} (x) - \bar{u} (x)$, the EMC and SMC data for
$g^p_1 (x)$, $g^n_1 (x)$ and $g^d_1 (x)$ \cite{EMC98},\cite{SMC98},
{\it without any free parameter} except for the starting energy scale of
the DGLAP evolution equation \cite{WK99},\cite{WW99}.
In the present paper, we have
evaluated the spin and orbital angular momentum distribution
functions within the same theoretical framework, and have
investigated their scale dependences.
It has been shown that the model can explain
the smallness of the quark spin fraction at least qualitatively
without assuming large gluon polarization at the low renormalization
point. A key ingredient here is the large and positive orbital
angular momentum carried by quarks and antiquarks with soft
(small $x$) component. The predicted large and positive quark
orbital angular momentum at the low renormalization point
necessarily affects the scenario of the scale dependence of the
nucleon spin contents. The solution of the LO evolution equation
indicates that the quark orbital angular momentum is a decreasing
function of $Q^2$ and it may play insignificant role at the
asymptotic energy scale, although the rate of reduction is
quite slow beyond $Q^2 \simeq 5 \,\mbox{GeV}^2$.

\vspace{10mm}
\noindent
\begin{large}
{\bf Acknowledgement}
\end{large}
\vspace{3mm}

We would like to express our thanks to A.~Sch\"{a}fer and
O.~Martin for providing us with their evolution program
for orbital angular momentum distributions and also for
many helpful discussions. We are also grateful to V. Vento
for clarifications on their results of Ref. \cite{SV99}.

%
%

\renewcommand{\baselinestretch}{1.0}

\end{document}